\documentclass[aps,prl,amsmath,amssymb,amsfonts,twocolumn,showpacs,superscriptaddress,groupedaddress]{revtex4}
\usepackage{amsthm}
\usepackage{epsfig}
\usepackage{graphicx}
\usepackage{subfigure}
\usepackage{dcolumn}
\usepackage{bm}

\newcommand{\del}{\partial}

\renewcommand{\Im}{\operatorname{Im}}

\begin{document}

\title{Action and entanglement in gravity and field theory}

\author{Yasha Neiman}
\email{yashula@gmail.com}
\affiliation{Institute for Gravitation \& the Cosmos and Physics Department, Penn State, University Park, PA 16802, USA}

\date{\today}

\begin{abstract}
In non-gravitational quantum field theory, the entanglement entropy across a surface depends on the short-distance regularization. Quantum gravity should not require such regularization, and it's been conjectured that the entanglement entropy there is always given by the black hole entropy formula evaluated on the entangling surface. We show that these statements have precise classical counterparts at the level of the action. Specifically, we point out that the action can have a non-additive imaginary part. In gravity, the latter is fixed by the black hole entropy formula, while in non-gravitating theories, it is arbitrary. From these classical facts, the entanglement entropy conjecture follows by heuristically applying the relation between actions and wavefunctions.
\end{abstract}

\pacs{03.65.Ud,04.20.Fy,04.70.Dy}  

\maketitle

\section{Introduction}

The Bekenstein-Hawking formula for black hole entropy \cite{Bekenstein:1972tm,Bekenstein:1973ur,Hawking:1974rv} is an important clue for any attempt at quantum gravity. In Planck units with $c = \hbar = 8\pi G = 1$, it relates the entropy $\sigma_H$ of a black hole with horizon $H$ to the horizon's area $A$ as:
\begin{align}
 \sigma_H = \frac{A}{4G} = 2\pi A \ . \label{eq:BH}
\end{align}
In classical diff-invariant theories of gravity other than General Relativity (GR), the entropy formula \eqref{eq:BH} is modified according to Wald's prescription \cite{Jacobson:1993xs,Wald:1993nt}. 

Despite much progress, the precise physical meaning and the range of applicability of the black hole entropy formula remains unclear. One possibility was recently articulated by Bianchi and Myers \cite{Bianchi:2012ev}. They conjecture that eq. \eqref{eq:BH} (generalized a-la Wald) is a universal formula for the entanglement entropy across a surface $H$, whenever the relevant states admit a semiclassical spacetime interpretation. Similar statements were made previously by many authors, as reviewed in \cite{Bianchi:2012ev}. This conjecture is in sharp contrast with the situation in non-gravitational quantum field theory. There, the entanglement entropy between adjoining regions is UV-divergent. Given a short-distance regulator, it will depend on the cutoff scale, as well as on the specific field theory. Thus, the Bianchi-Myers conjecture implies that gravity provides a universal short-distance regulator (at the Planck scale) for all possible sets of matter fields.

Strictly speaking, entanglement entropy is only defined when the total state of the system is pure. This will not be the case in general, especially if the total state is itself defined in a bounded region of space. We therefore consider a slight generalization of the Bianchi-Myers conjecture. Let there be a semiclassical state in a spatial region $AB$, separated into subregions $A$ and $B$ by a surface $H$. We can then state the conjecture, along with the UV-sensitivity of entanglement in non-gravitating theories, as:
\begin{align}
 \sigma_A + \sigma_B - \sigma_{AB} = \left\{
  \begin{array}{ll}
    2\sigma_H & \quad \text{(gravity)} \\
    \text{anything} & \quad \text{(no gravity)}
  \end{array} \right. \ . \label{eq:mutual}
\end{align}
Here, $\sigma_A$, $\sigma_B$ and $\sigma_{AB}$ are the von Neumann entropies of the state in the corresponding regions, and $\sigma_H$ is the black hole entropy formula evaluated on $H$. The LHS of \eqref{eq:mutual} is known as the mutual information. When the overall state is pure, we have $\sigma_{AB} = 0$, while $\sigma_A$ and $\sigma_B$ both equal the entanglement entropy. This is the reason for the factor of 2 on the RHS. 

\section{Summary of results}

\begin{figure*}%
\centering%
\includegraphics[scale=.9]{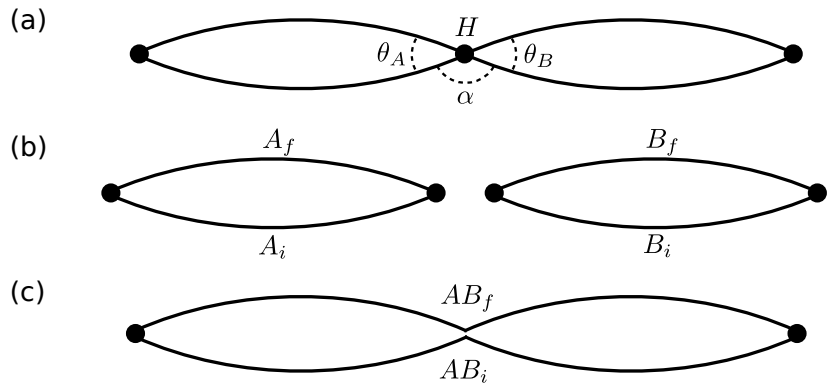} \\
\caption{A spacetime process (a) joined along an ``entangling surface'' $H$. The process can be viewed either (b) as two separate evolutions of the spatial regions $A$ and $B$, or (c) as a single evolution of the overall spatial region $AB$.}
\label{fig:entangle} 
\end{figure*}

In this paper, we point out that eq. \eqref{eq:mutual} has a classical counterpart in the actions of adjoining spacetime processes. Consider a spacetime region of the type depicted in figure \ref{fig:entangle}(a). It describes an evolution between initial and final states on spacelike hypersurfaces, joined together along an ``entangling surface'' $H$. This can be viewed in two different ways, illustrated in figures \ref{fig:entangle}(b,c). In one view, there are two separate causally closed processes, taking place in each of the spatial regions $A$ and $B$, with respective actions $S_A$ and $S_B$. In the other view, there is a single process taking place in the larger region $AB$ with action $S_{AB}$, such that the spacelike hypersurfaces housing the initial and final states happen to intersect at the surface $H$. 

Our central statement is as follows. In a classical theory of gravity that is second-order in time derivatives, the actions $S_A$, $S_B$ and $S_{AB}$ satisfy the relation:
\begin{align}
 S_A + S_B - S_{AB} = \left\{
  \begin{array}{ll}
    i\sigma_H & \quad \text{(gravity)} \\
    i\cdot\text{anything} & \quad \text{(no gravity)}
  \end{array} \right. \, , \label{eq:result}
\end{align}
where $\sigma_H$ is again the black hole entropy formula evaluated on $H$. Eq. \eqref{eq:result} encompasses several properties of the action that will be surprising for many readers. First, as noticed by Brill and Hayward \cite{Brill:1994mb}, the gravitational action is non-additive. Second, as noticed by the author \cite{Neiman:2013ap,Bodendorfer:2013hla,Neiman:2013lxa}, it has an imaginary part that is closely related to the black hole entropy formula. Third, in non-gravitational field theory, these properties do not necessarily disappear, but instead their magnitude becomes undetermined. The above features are all related to the corner contributions to the action's boundary term. The necessary details will be reviewed and developed in the sections that follow.

The similarity between eqs. \eqref{eq:mutual} and \eqref{eq:result} seems more than just superficial. In fact, on a heuristic level, the formula \eqref{eq:result} for the action's non-additivity \emph{implies} the mutual information conjecture \eqref{eq:mutual}. The reasoning is as follows. First, action differences exponentiate into wavefunction ratios: 
\begin{align}
 \frac{\psi_A\psi_B}{\psi_{AB}} = e^{i(S_A + S_B - S_{AB})} \ .  
\end{align}
These square into ratios of density matrix eigenvalues: 
\begin{align}
 \frac{\rho_A\rho_B}{\rho_{AB}} = \frac{\left|\psi_A\psi_B\right|^2}{\left|\psi_{AB}\right|^2} = e^{-2\Im(S_A + S_B - S_{AB})} \ .  
\end{align}
Finally, the logarithms of density matrix eigenvalues give entropies, providing the desired link from \eqref{eq:result} to \eqref{eq:mutual}: 
\begin{align}
 \begin{split}
  \sigma_A + \sigma_B - \sigma_{AB} &= -\langle\ln\rho_A\rangle - \langle\ln\rho_A\rangle + \langle\ln\rho_{AB}\rangle \\ 
    &= 2\Im(S_A + S_B - S_{AB}) \ .
 \end{split}
\end{align}
These considerations suggest a new role for the classical action: through its non-additive imaginary part, the action knows about the entanglement entropy of semiclassical states. However, as stated in \eqref{eq:result} and will be shown below, this can only be put to good use in gravitational theories: otherwise, there is too much freedom in the action's definition.

\section{Non-additivity of the gravity action} \label{sec:add}

In this section, we review the relevant facts from \cite{Brill:1994mb} on the non-additivity of the GR action due to corner contributions. We postpone the discussion of the action's imaginary part by considering first the Euclidean theory, where the action is real. While we consider GR for simplicity, the results extend straightforwardly to Lovelock gravity \cite{Lovelock:1971yv}, using the corner contributions from \cite{Neiman:2013ap} and their behavior under $2\pi$ rotations from \cite{Banados:1993qp,Neiman:2013lxa}.     

The action of GR in a spacetime region $\Omega$ with boundary $\del\Omega$ is given by: 
\begin{align}
 \begin{split}
  S ={}& \int_\Omega \sqrt{\pm g}\left(-\frac{1}{2}R \pm \Lambda + \mathcal{L}_M \right) d^dx \\
   &+ \int_{\del\Omega} \sqrt{\frac{\pm h}{n\cdot n}}\,(-K + C)\, d^{d-1}x \ . \label{eq:S_GR}
 \end{split}
\end{align}
Here, the $\pm$ signs correspond to Euclidean and Lorentzian spacetime, respectively. $g_{\mu\nu}$ is the spacetime metric (with mostly-minus signature in the Lorentzian), $g$ and $R$ are its determinant and Ricci scalar, $\Lambda$ is the cosmological constant, and $\mathcal{L}_M$ is the (minimally coupled) matter Lagrangian. In the boundary term \cite{York:1972sj,Gibbons:1976ue}, $h_{ab}$ is the intrinsic metric, $h$ is its determinant, $n_\mu$ is the boundary normal oriented such that $n_\mu v^\mu > 0$ for outgoing vectors $v^\mu$, and $K = \nabla_a n^a$ is the trace of the extrinsic curvature. $C$ is an arbitrary functional of $h_{ab}$ and matter fields that doesn't contain normal derivatives.

Consider now a region shaped as in figure \ref{fig:entangle}, but in Euclidean spacetime. At the surface $H$, the boundary has a corner: it is non-differentiable, and turns suddenly by a finite angle. At such surfaces, the extrinsic curvature has a delta-function singularity. When the $K$ term in \eqref{eq:S_GR} is integrated through this singularity, it picks up a ``corner contribution'' of $A(\theta - \pi)$ \cite{Hartle:1981cf,Hayward:1993my}, where $A$ is the corner's area, and $\theta$ is the dihedral angle between the intersecting hypersurfaces. 

The crucial point is that the $\pi A$ pieces in the corner contributions are non-additive under the gluing together of spacetime regions. In our setup, this plays out as follows. Let $\theta_{A,B}$ be the dihedral angles indicated in figure \ref{fig:entangle}(a). Then the corner contributions at $H$ in figure \ref{fig:entangle}(b) read:
\begin{align}
 A(\theta_A - \pi) + A(\theta_B - \pi) = A(\theta_A + \theta_B - 2\pi) \ . \label{eq:corner_separate}
\end{align}
After the gluing, i.e. in figure \ref{fig:entangle}(c), the relevant dihedral angles are instead $2\pi - \alpha$ and $\theta_A + \theta_B + \alpha$, with $\alpha$ as indicated in figure \ref{fig:entangle}(a). The corner contributions at $H$ become:
\begin{align}
 A(\pi - \alpha) + A(\theta_A + \theta_B + \alpha - \pi) = A(\theta_A + \theta_B) \ . \label{eq:corner_glued}
\end{align}
The other terms in the action \eqref{eq:S_GR} are additive. Though the $C$ term may have singularities at $H$ due to discontinuities in the derivatives of $h_{ab}$, the transition from figure \ref{fig:entangle}(b) to \ref{fig:entangle}(c) only changes the order in which these are integrated. Thus, the non-additivity of the action comes entirely from the extrinsic curvature term, and reads:
\begin{align}
 S_A + S_B - S_{AB} = -2\pi A = -\sigma_H \ . \label{eq:Euclid_result}
\end{align}
In higher-curvature Lovelock gravity, the corner contributions \cite{Neiman:2013ap} are more complicated, involving integrals over the angle of $n^\mu$. The difference $S_A + S_B - S_{AB}$ can then be expressed as an integral over a full turn. From the analysis in \cite{Banados:1993qp,Neiman:2013lxa}, one again obtains the relation \eqref{eq:Euclid_result} with the black hole entropy formula $\sigma_H$ (but not with the surface area $A$).    

\section{Imaginary part of the gravity action} \label{sec:imaginary}

\begin{figure}%
\centering%
\includegraphics[scale=0.65]{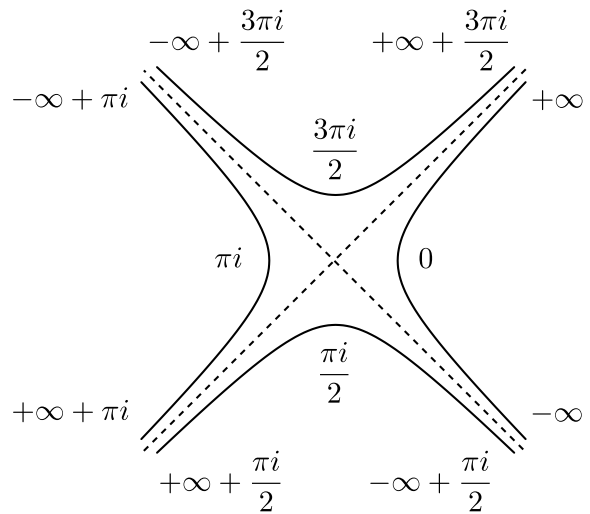} \\
\caption{An assignment of boost angles in the Lorentzian plane. The horizontal and vertical axes describe the spacelike and timelike components of a vector $n^\mu$. The sign choices for the real and imaginary parts of the angle are separate. The angles are defined up to integer multiples of $2\pi i$.}
\label{fig:angles_plane} 
\end{figure}

We now return to Lorentzian GR. For rotation angles around timelike corner surfaces, the situation is the same as in the Euclidean, up to signs. However, in figure \ref{fig:entangle}, the surface $H$ is spacelike, while the dihedral angles around it are boosts in a Lorentzian plane. Normally, one considers boost angles within a single quadrant of the plane, where they span the entire real range $(-\infty,\infty)$. However, this is not enough for the corners in figure \ref{fig:entangle}(b): the boundary normal there goes from past-pointing timelike on the initial hypersurface to future-pointing timelike on the final one. This requires crossing two quadrant boundaries in the Lorentzian plane. To evaluate such corner contributions, we must assign boost angles to the entire plane, rather than just to a single quadrant. This assignment is illustrated in figure \ref{fig:angles_plane}. It was found in \cite{SorkinThesis} by analytically continuing the inverse trigonometric functions, and was rederived in \cite{Neiman:2013ap} using a contour integral. The crucial point is that in order to span the entire plane, the angle must become complex. Specifically, the angle picks up an imaginary contribution $\pi i/2$ per quadrant crossing, i.e. per signature flip of the rotating vector. The angle for a full turn is then $2\pi i$, in analogy with the Euclidean $2\pi$. In the contour-integral approach, the $\pi i/2$ jumps come from bypassing poles in a $dz/z$ integral, after decomposing the rotating vector as $n^\mu \sim L^\mu + z\ell^\mu$ in a null basis $(L^\mu,\ell^\mu)$.

The imaginary parts of corner angles plug into the action's corner contributions, making the action complex. While the sign of the angles' real part is determined by the sign of the boundary term in \eqref{eq:S_GR}, one must make a separate choice for the sign of the imaginary part. This reflects the two separate $P,T$ transformations that are present in Lorentzian signature. The choice argued for in \cite{Neiman:2013ap} is the one that makes $\Im S$ positive. This leads to amplitudes $e^{iS}$ that are exponentially damped rather than exploding. 

In analogy with \eqref{eq:corner_separate}, we now have the corner contributions from $H$ in figure \ref{fig:entangle}(b) as:
\begin{align}
 A(\pi i - \theta_A) + A(\pi i - \theta_B) = A(2\pi i - \theta_A - \theta_B) \ ,
\end{align}
where $\theta_{A,B}$ are the (real, positive) dihedral angles from figure \ref{fig:entangle}(a). For figure \ref{fig:entangle}(c), the relevant dihedral angles are now $2\pi i - \alpha$ and $\theta_A + \theta_B + \alpha$, with $\Im\alpha = \pi$. The corresponding corner contributions to the action read:
\begin{align}
 A(\alpha - \pi i) + A(\pi i - \theta_A - \theta_B - \alpha) = -A(\theta_A + \theta_B) \ .
\end{align}
Thus, the action's non-additivity takes the form:
\begin{align}
 S_A + S_B - S_{AB} = 2\pi i A = i\sigma_H \ , \label{eq:result_derived}
\end{align}
which establishes the upper line in eq. \eqref{eq:result}. The analogous result in Lovelock gravity again follows similarly, using the results of \cite{Neiman:2013ap,Neiman:2013lxa} for the imaginary parts of the corner integrals. We expect the result to also hold for arbitrary two-time-derivative matter couplings. For example, a conformally coupled scalar is handled straightforwardly, as discussed in \cite{Neiman:2013lxa}.

\section{The ambiguity in non-gravitational actions} \label{sec:field_theory}

It remains to establish the lower line in eq. \eqref{eq:result}, i.e. that non-gravitational field theory actions have an \emph{arbitrary} non-additive imaginary part. This is surprising at first, since the matter Lagrangian $\mathcal{L}_M$ in \eqref{eq:S_GR} has no such property. The crucial point is that when the metric is non-dynamical, we have a lot of freedom in redefining the action without disturbing its variational principle. For example, we can add to the Lagrangian any functional of the metric. More to the point, we can add to the boundary term any functional of the metric $h_{ab}$, the extrinsic curvature $K_{ab}$ and the matter fields (but not the matter fields' normal derivatives). In particular, we can add any multiple of the GR boundary term or its Lovelock-gravity generalizations \cite{Myers:1987yn}. In this way, we can write different actions that vary identically under variations of the matter fields, while having arbitrary non-additive imaginary parts. In gravitational theories, this freedom is not present. There, the extrinsic curvature term is fixed by the Lagrangian, so as to satisfy a variational principle where $h_{ab}$ but not $K_{ab}$ is held fixed on $\del\Omega$.

The ambiguity in the non-gravitational action is not just an esoteric loophole. It's manifested whenever we take the limit from a gravitating to a non-gravitating theory by sending the coupling $G$ to zero. Consider again the GR action \eqref{eq:S_GR}, setting $\Lambda=C=0$ and restoring the units of $G$:
\begin{align}
 \begin{split}
  S ={}& \int_\Omega \sqrt{-g}\,\mathcal{L}_M d^dx - \frac{1}{16\pi G}\int_\Omega \sqrt{-g}\,R\, d^dx \\
   &- \frac{1}{8\pi G}\int_{\del\Omega} \sqrt{\frac{-h}{n\cdot n}}\,K\, d^{d-1}x \ . \label{eq:S_ambig}
 \end{split} 
\end{align}
As we send $G$ to zero, this action does not reduce to the $\mathcal{L}_M$ term! By the Einstein equations, the bulk Einstein-Hilbert term has a finite limit:
\begin{align}
 \frac{1}{16\pi G}\,R = \frac{1}{2-d}\,T_\mu^\mu \ ,
\end{align}
where $T_\mu^\nu$ is the matter stress-energy tensor. As for the York-Gibbons-Hawking boundary term, it diverges: the boundary's extrinsic curvature does not become small as $G$ is sent to zero. In particular, the non-additive imaginary part \eqref{eq:result_derived} diverges as:
\begin{align}
 S_A + S_B - S_{AB} = \frac{i}{4G}\,A \ \rightarrow\ i\infty \ .
\end{align}
This is consistent with the expectation that the mutual information diverges as the short-distance cutoff $G$ is removed.

\section{Discussion} \label{sec:discuss}

We have shown that the action in gravitational and non-gravitational field theories has a property \eqref{eq:result} that is directly analogous to the mutual information conjecture \eqref{eq:mutual}. The gravitational part of the result is demonstrated for GR with minimally coupled matter and for Lovelock gravity, and probably holds for all two-time-derivative theories. The restriction to two time derivatives is necessary to ensure a standard variational principle, and thus a well-motivated boundary term. Put differently, it implies that the variational principle makes sense at arbitrarily short distance scales. This requirement is consistent with the fact that both the mutual information and the action's corner contributions are short-distance effects.

We've seen that the conjecture \eqref{eq:mutual} can be heuristically \emph{derived} from the action's property \eqref{eq:result}. This suggests that the Lorentzian classical action plays a role in encoding state statistics, in addition to its role in the variational principle (as discussed in \cite{Neiman:2013ap}, the variational principle is unaffected by the action's imaginary part). These issues should be explored further. A possible avenue may be the general-boundary approach to mixed and entangled states, along the lines of \cite{Bianchi:2013toa}.

\section*{Acknowledgements}		

I am grateful to Eugenio Bianchi, Norbert Bodendorfer and Rob Myers for discussions. This work is supported in part by the NSF grant PHY-1205388 and the Eberly Research Funds of Penn State.


\begin{thebibliography} {99}

\bibitem{Bekenstein:1972tm} 
  J.~D.~Bekenstein,
  Lett.\ Nuovo Cim.\  {\bf 4}, 737 (1972).

\bibitem{Bekenstein:1973ur} 
  J.~D.~Bekenstein,
  Phys.\ Rev.\ D {\bf 7}, 2333 (1973).

\bibitem{Hawking:1974rv} 
  S.~W.~Hawking,
  Nature {\bf 248}, 30 (1974).

\bibitem{Jacobson:1993xs} 
  T.~Jacobson and R.~C.~Myers,
  Phys.\ Rev.\ Lett.\  {\bf 70}, 3684 (1993).

\bibitem{Wald:1993nt} 
  R.~M.~Wald,
  Phys.\ Rev.\ D {\bf 48}, 3427 (1993).

\bibitem{Bianchi:2012ev} 
  E.~Bianchi and R.~C.~Myers,
  arXiv:1212.5183 [hep-th].

\bibitem{Brill:1994mb} 
  D.~Brill and G.~Hayward,
  Phys.\ Rev.\ D {\bf 50}, 4914 (1994).

\bibitem{Neiman:2013ap} 
  Y.~Neiman,
  JHEP {\bf 1304}, 071 (2013).

\bibitem{Bodendorfer:2013hla} 
  N.~Bodendorfer and Y.~Neiman,
  Class.\ Quant.\ Grav.\  {\bf 30}, 195018 (2013).

\bibitem{Neiman:2013lxa} 
  Y.~Neiman,
  Phys.\ Rev.\ D {\bf 88}, 024037 (2013).

\bibitem{Lovelock:1971yv} 
  D.~Lovelock,
  J.\ Math.\ Phys.\  {\bf 12}, 498 (1971).

\bibitem{Banados:1993qp} 
  M.~Banados, C.~Teitelboim and J.~Zanelli,
  Phys.\ Rev.\ Lett.\  {\bf 72}, 957 (1994).

\bibitem{York:1972sj} 
  J.~W.~York, Jr.,
  Phys.\ Rev.\ Lett.\  {\bf 28}, 1082 (1972).

\bibitem{Gibbons:1976ue} 
  G.~W.~Gibbons and S.~W.~Hawking,
  Phys.\ Rev.\ D {\bf 15}, 2752 (1977).

\bibitem{Hartle:1981cf} 
  J.~B.~Hartle and R.~Sorkin,
  Gen.\ Rel.\ Grav.\  {\bf 13}, 541 (1981).

\bibitem{Hayward:1993my} 
  G.~Hayward,
  Phys.\ Rev.\ D {\bf 47}, 3275 (1993).

\bibitem{SorkinThesis}
  R.~Sorkin,
  Ph.D. thesis, California Institute of Technology, 1974. 

\bibitem{Myers:1987yn} 
  R.~C.~Myers,
  Phys.\ Rev.\ D {\bf 36}, 392 (1987).

\bibitem{Bianchi:2013toa} 
  E.~Bianchi, H.~M.~Haggard and C.~Rovelli,
  arXiv:1306.5206 [gr-qc].

\end{thebibliography}
\end{document}